\documentclass[trackchanges,twocolumn]{aastex7}

\hypersetup{linkcolor=red,citecolor=blue,filecolor=cyan,urlcolor=blue}

\usepackage{graphicx,color}
\usepackage{amssymb}
\usepackage{amsmath}
\usepackage{url}
\usepackage{natbib}
\usepackage{txfonts}
\usepackage{multirow}
\usepackage{array}
\usepackage{rotating}
\usepackage{sidecap}
\usepackage{hyperref}
\usepackage{epstopdf}
\usepackage{footnote}
\usepackage{tabularx}
\usepackage{booktabs}
\usepackage{longtable}
\usepackage{tabu}
\usepackage{longtable}

\newcommand{\kms}{km~s$^{-1}$}

\newcommand{\sdo}{\textit{SDO}}

\begin{document}

\title  {H$\beta$ Spicules, Small-scale Jets, and H$\beta$ Microflashes: Sub-arcsecond Dynamic Events and their Magnetic Origins in the Lower Solar Chromosphere Observed by DKIST}

\author[0000-0001-7620-362X]{Navdeep K. Panesar}
\affiliation{Lockheed Martin Solar and Astrophysics Laboratory, 3251 Hanover Street, Bldg. 203, Palo Alto, CA 94306, USA}
\affiliation{SETI Institute, 339 Bernardo Ave, Mountain View, CA 94043, USA}
\email{panesar@lmsal.com}

\author[0000-0002-0824-3109]{V. Aparna}
\affil{Department of Astronomy, New Mexico State University, Las Cruces, NM 88003, USA. }
\email{aparna@baeri.org}
\author[0000-0001-7817-2978]{Sanjiv K. Tiwari}
\affil{Lockheed Martin Solar and Astrophysics Laboratory, 3251 Hanover Street, Bldg. 203, Palo Alto, CA 94306, USA}
\affiliation{SETI Institute, 339 Bernardo Ave, Mountain View, CA 94043, USA}
\affil{Bay Area Environmental Research Institute, NASA Research Park, Moffett Field, CA 94035, USA}
\email{tiwari@lmsal.com}
\author[0000-0003-1281-897X]{Alphonse C. Sterling}
\affil{NASA Marshall Space Flight Center, Huntsville, AL 35812, USA}
\email{alphonse.sterling@nasa.gov}

\author[0000-0002-5691-6152]{Ronald L. Moore}
\affil{NASA Marshall Space Flight Center, Huntsville, AL 35812, USA}
\affil{Center for Space Plasma and Aeronomic Research (CSPAR), UAH, Huntsville, AL 35805, USA}
\email{ronald.l.moore@nasa.gov}
\author[0000-0002-3009-295X]{J. M. da Silva Santos}
\affiliation{National Solar Observatory, 3665 Discovery Drive, Boulder, CO 80303, USA}
\email{jdasilvasantos@nso.edu}

\begin{abstract}

 We examine an on-disk solar quiet region of enhanced magnetic network 
 using H$\beta$  images from Daniel K. Inouye Solar Telescope's (DKIST's) Visible Broadband Imager (VBI) and line-of-sight magnetograms from DKIST’s Visible Spectro-Polarimeter (ViSP). We also compare the ViSP magnetograms with co-aligned co-temporal line-of-sight magnetograms from the Solar Dynamics Observatory (SDO)/Helioseismic and Magnetic Imager (HMI). We find:
 (i) Two types of  chromospheric jet-like features: H$\beta$ spicules, and small-scale jets. Both are rooted near edges of magnetic network lanes. Several sit close to ViSP-detected tiny islands of either minority-polarity flux or dips in majority-polarity flux. Hence, those several plausibly stem from mixed-polarity magnetic flux, and are caused by chromospheric magnetic reconnection. For H$\beta$ spicules, the average width, length, lifetimes, and speeds are 420$\pm$400 km, 2600$\pm$1600 km, 4.3$\pm$0.25 min, and 12$\pm$4.7 \kms. For small-scale jets, those are 385$\pm$100 km, 620$\pm$40 km, 5$\pm$3min, and 4.4$\pm$1.5 \kms. (ii) H$\beta$ microflashes sit in evidently unipolar flux. Their widths, lengths, lifetimes, and speeds are 180$\pm$65 km, 365$\pm$100 km, 4.3$\pm$2.6 min, and 3.1$\pm$0.2 \kms. (iii) At the base of a coronal plume, due to its higher spatial resolution than HMI, ViSP shows both some very strong ($>$800 G) majority-polarity flux not shown by HMI and some sub-arc-second minority-polarity-flux inclusions not shown by HMI. 
 These sub-arcsecond chromospheric transients carry sufficient energy ($\simeq$10$^{24}$erg) to transiently heat the local chromosphere and corona and potentially contribute to solar-wind acceleration. They may represent the small-scale end of a continuum of magnetic-reconnection-driven activity, highlighting the importance of DKIST's high-resolution magnetic-field measurements for understanding small-scale chromospheric dynamics.

\end{abstract}


\section{Introduction}

The solar chromosphere exhibits a wide variety of fine-scale, highly dynamic structures that reflect the complex interaction between plasma motions and magnetic fields. Prominent among these are fibrils, spicules, mottles, and jets, which are ubiquitously observed in both plage and magnetic network regions \citep{beckers68,sterling00b,tsiropoula2012}. These features are closely aligned with the local magnetic field and are widely interpreted as tracers of chromospheric magnetic topology \citep{rodrguez2011}. Observational and numerical studies further suggest that their formation and dynamics are governed by a combination of magnetoacoustic waves that steepen into shocks and small-scale magnetic reconnection processes, highlighting their role in chromospheric energy and mass transport \citep{hansteen06,pontieu07-spicules,shibata07}.

\begin{figure*}[ht!]
	\centering
	\includegraphics[width=\linewidth]{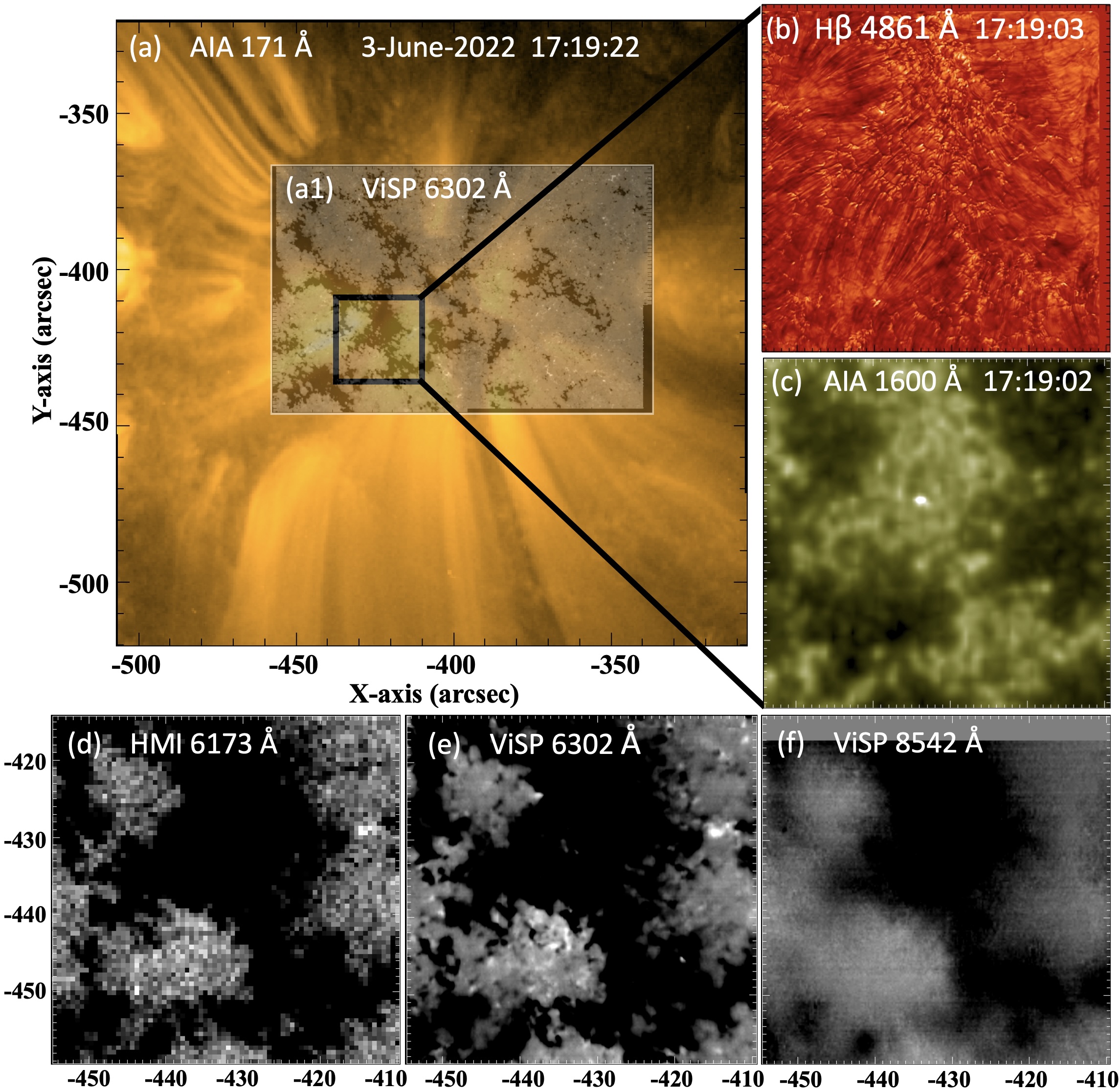}
	\caption{The observed quiet region. Panel (a) is an SDO/AIA 171 \AA\  image of the region.  The black box outlines the field of view (FOV) in co-temporal panels (b – f).  Inset (a1) shows the full FOV of a co-temporal ViSP line-of-sight magnetogram.  Panel (b) is a VBI H$\beta$ (4861 \AA) image.  Panel (c) is an AIA 1600 \AA\  image.  Panel (d) is an SDO/HMI LOS magnetogram constructed from ViSP slit positions.  Panels (e) and (f) are LOS magnetograms of the flux measured from the ViSP $\lambda$6302 and $\lambda$8542 lines, respectively.
		\label{fig1}}
\end{figure*}

High-spatial-resolution H$\alpha$ observations have revealed a rich variety of small-scale dynamic features in plage regions \citep{berger99,pontieu99}. Together with morphologically similar structures observed elsewhere in the chromosphere, these features have been described using terms such as dynamic fibrils \citep[e.g.][]{pontieu07a-fibrils} and Type I spicules \citep[e.g.][]{pontieu07-spicules}. Many of these H$\alpha$ features are believed to originate from wave-driven oscillations that steepen into shocks as they propagate through the chromosphere \citep{hansteen06}. They also exhibit strong similarities to Ca II chromospheric anemone jets \citep{shibata07}, which have been proposed to result from magnetic reconnection between emerging or pre-existing bipolar magnetic fields and the surrounding ambient field. In addition, some of these features resemble coronal jets that exhibit base brightenings at the onset of the eruption \citep{shimojo96}. Such jets are thought to be driven by the eruption of minifilaments \citep{sterling15}, which are triggered by magnetic flux cancelation inside the jet base \citep{panesar16b,panesar17,panesar18a,mcglasson19}. Collectively, these small-scale transient phenomena may contribute to coronal heating and solar-wind acceleration by transporting energy and mass into the upper solar atmosphere \citep{pontieu04,innes09,moore11,jeongwoo24,sterling2024,panesar2026}.

Here we present a study of sub-arcsecond chromospheric dynamic events (namely H$\beta$ spicules, small-scale jets, and H$\beta$ microflashes) seen in the exceptionally high-resolution images and magnetograms from the Daniel K. Inouye Solar Telescope (DKIST; \citealt{rimmele20}) and compare the DKIST observations with co-temporal observations from Solar Dynamics Observatory (SDO; \citealt{pesnell12}).  

Our H$\beta$ spicules correspond to what have long been called solar spicules or mottles.  In the following, we in particular call out the features that we refer to as H$\beta$ spicules as having properties similar to spicules.  We note that these objects we observe here are extremely highly resolved, and may not always take on the appearance of traditional H$\alpha$ spicules, making an accurate and informed comparison difficult.  Additionally, high-resolution H$\beta$ spicule images are not common \citep{krat1971,alissandrakis1973,bose23}, with most images being taken in H$\alpha$, and more recently in Ca II.  It is likely that we are observing elements of spicules,  and strands or bundles of strands of spicules.  Please see an extensive debate in the literature for the types and names of various spicule-like features \citep[e.g.,][]{pontieu07-spicules, pereira12, zhang-spicule12, panesar19, sterling-spicule2021}.

DKIST's high spatial resolution enables detection of fine-scale photospheric magnetic flux clumps not seen by HMI.  In addition to characterizing chromospheric fine-scale short-lived structure, we examine their magnetic origins and investigate whether there is fine-scale minority-polarity flux that is at or near their base but is undetected in HMI magnetograms.

\begin{figure*}[ht!]
	\centering
	\includegraphics[width=0.9\linewidth]{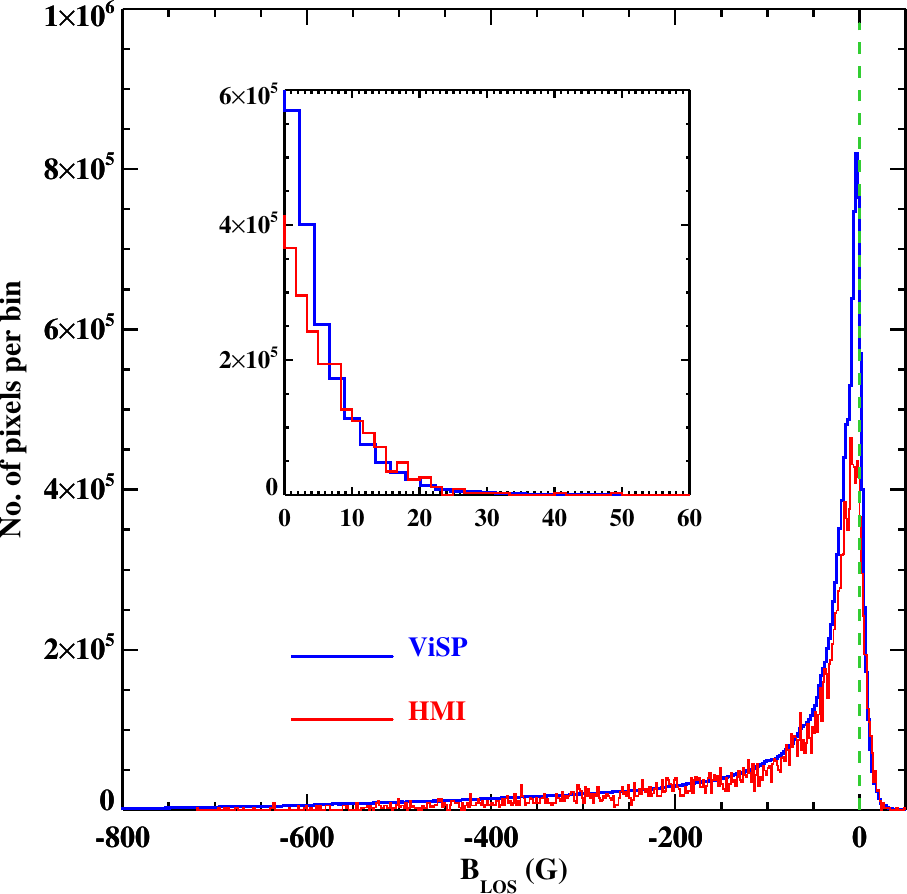}
	\caption{Comparison of ViSP and HMI magnetic flux. The histograms are of the magnetic flux in the pixels covering the analyzed FOV for ViSP and HMI, the FOV of Figure \ref{fig1}d and Figure \ref{fig1}e. The ViSP pixel-flux histogram is blue; the HMI pixel-flux histogram is red. The green dashed vertical line marks 0 G. The inset for the positive flux shows  the enhanced sensitivity of ViSP to weaker flux compared to HMI. 
		\label{fig2}}
\end{figure*}

\begin{figure*}
	\centering
	\includegraphics[width=0.7\linewidth]{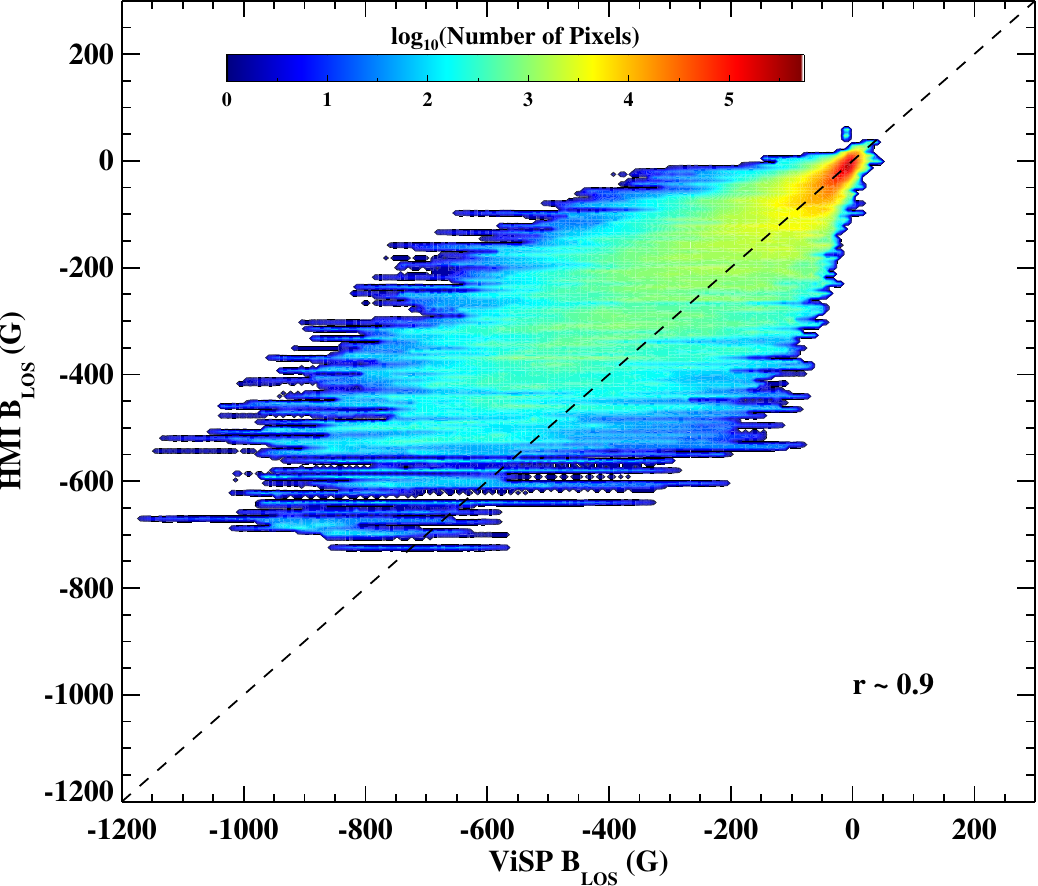}
	\caption{A density plot of the ViSP and HMI line-of-sight magnetic field measurements for the analyzed FOV shown in Figures \ref{fig1}d and \ref{fig1}e. The color scale represents $\log_{10}$(Number of Pixels). The black dashed line denotes the 1:1 reference line, and $r$ indicates the Pearson correlation coefficient between the ViSP and HMI measurements.  
		\label{fig2a}}
\end{figure*}

\begin{figure*}
	\centering
	\includegraphics[width=\linewidth]{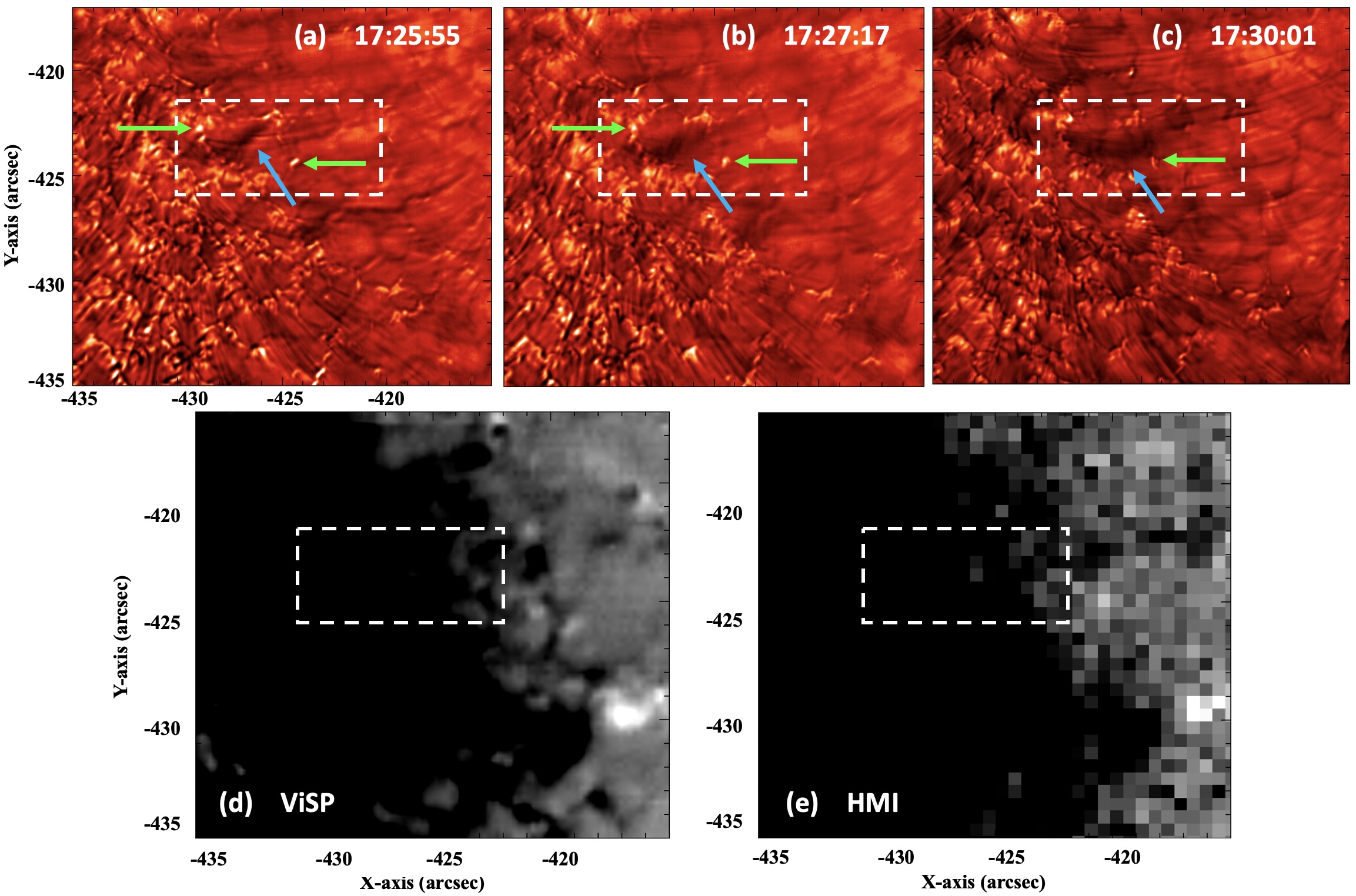}
	\caption{ Example H$\beta$ spicule and two example microflashes.  Panels (a – c) show VBI H$\beta$ images of a spicule (blue arrow) and two H$\beta$ microflashes (green arrows).  Panels (d) and (e) show a ViSP magnetogram and a synchronic HMI magnetogram of the same area.  The synchronic HMI magnetogram is constructed from the spatiotemporal locations of the ViSP slit.  Magnetic flux values are displayed within a range of $\pm$30 G. The white box outlines the area where the spicule and microflashes occur. An animation of panels (a) - (c) is available. This animation shows a larger field of view and runs from June 3rd, 2022 at 17:10:50 to 17:35:30. The animation real-time duration is 1.9 seconds. The same white dashed box and blue and green arrows are used in the animation.
		\label{fig3}}
\end{figure*}

\section{Observations, Data, and Methods} \label{sec:observations}

In this study, we analyze high-resolution observations of an enhanced network quiet region from DKIST’s Visible Broadband Imager (VBI; \citealt{woger21-vbi}) and DKIST’s Visible Spectro-Polarimeter (ViSP; \citealt{wijn22-visp}). The VBI H$\beta$ filtergrams provide a relatively large field of view (FOV) while resolving fine-scale chromospheric structures at sub-arcsecond resolution, enabling detailed analysis of their morphology and dynamics. Complementary ViSP spectropolarimetric measurements in the Fe I 6301/6302 \AA\ lines give the net magnetic flux in each pixel. These combined datasets allow us to investigate the connection between the magnetic field and the properties of chromospheric fibrils, as well as to study small-scale dynamic events in both the photosphere and chromosphere with high spatial resolution (0.0106\arcsec\ pixel$^{-1}$). 

We used and analyzed Level 1 DKIST observations of a network region from 2022 June 3; this dataset was publicly released in 2023 April. The ViSP instrument employed two of its three spectrograph arms to observe the Fe I 6301/6302 \AA\ and Ca II 8542 \AA\ spectral lines. The observations consisted of eight raster scans, each comprising 490 steps and requiring approximately 27 minutes to complete. The eight scans correspond to two repetitions of a four-tile mosaic, covering a total field of view of 210\arcsec\ $\times$ 136\arcsec\ for the 6302 \AA\ arm and 210\arcsec\ $\times$  100\arcsec\ for the 8542 \AA\ arm. The resulting mosaic cadence was approximately 113 minutes. For more details on this dataset, see papers  by \cite{santos23}, and \cite{kuridze24}.  The VBI data  consist of speckle-reconstructed, high-resolution  images acquired in the G-band, Ca II K, and H$\beta$ filters. Each ViSP raster scan is accompanied by mosaics from four different VBI co-pointings. Each of these four VBI maps is composed of nine subfields, together covering a full field of view of 115\arcsec\ $\times$ 117\arcsec\ (examples are shown in Figure 1 of \citealt{santos23}). Here in this paper, we focus on one VBI pointing (Figure \ref{fig1}) that is covered largely by enhanced magnetic network, and the time cadence between VBI frames is 73 to 96 seconds.  We crop the region corresponding to the central frame of the VBI full-FOV in the ViSP reconstructed magnetogram. 

\begin{figure*}
	\centering
	\includegraphics[width=\linewidth]{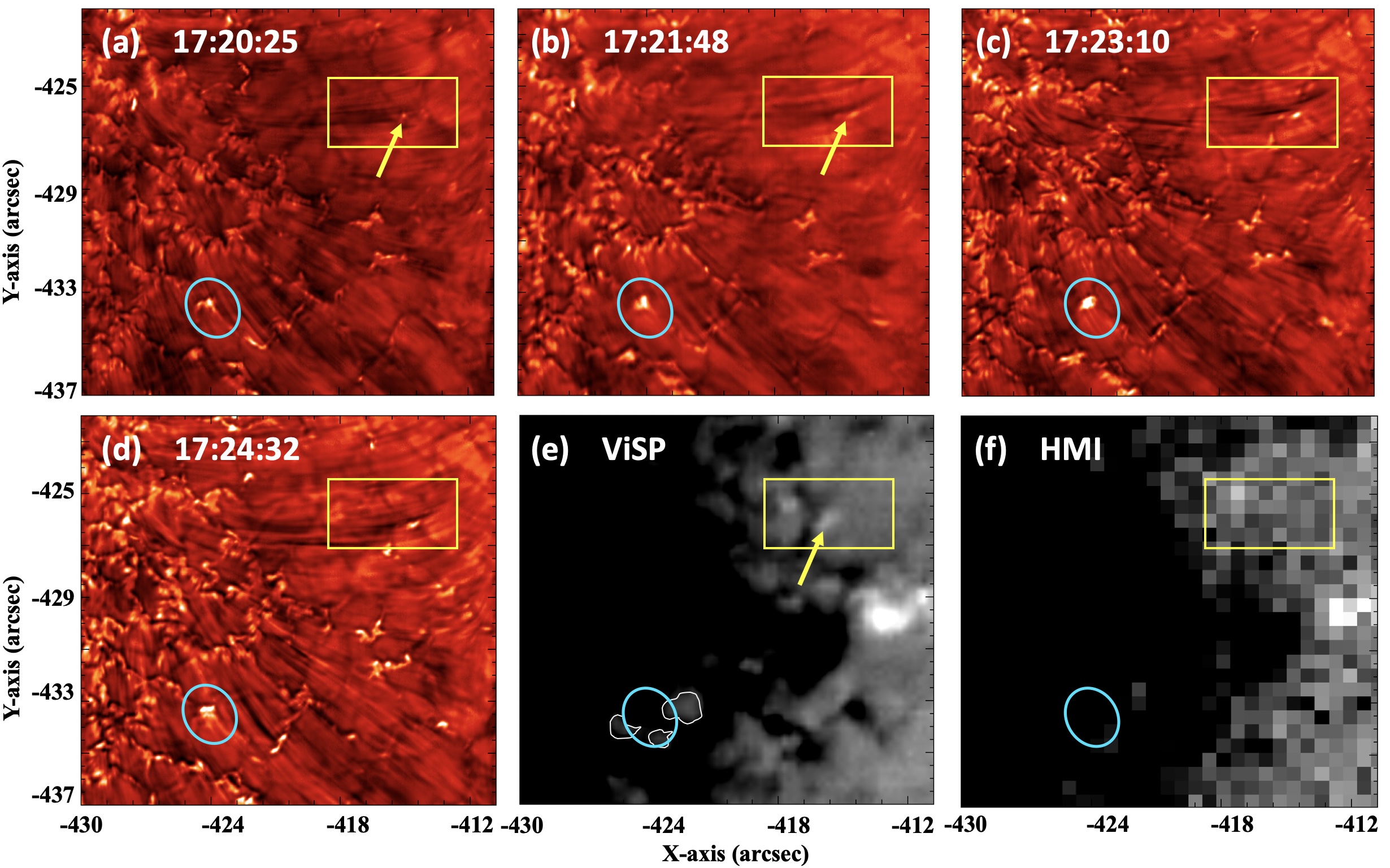}
	\caption{Example H$\beta$ spicule and example small-scale jet.  Panels (a – d) show VBI H$\beta$ images showing an example spicule (yellow arrows) and an example small-scale jet (blue oval).  Panels (e) and (f) present the corresponding ViSP and HMI magnetograms of the same region. The white contours in (e) outline the magnetic flux dips near the jet site. Magnetic field values are displayed within a range of $\pm$30 G. The yellow box outlines the location of the H$\beta$ spicule and the blue oval encloses the jet location. An animation of panels (a) - (d) is available. This animation shows a larger field of view and runs from June 3rd, 2022 at 17:10:50 to 17:35:30. The animation real-time duration is 1.9 seconds. The same yellow box and arrows plus blue ovals are used in the animation. 
		\label{fig4}}
\end{figure*}

We use co-temporal observations from SDO/AIA \citep{lem12}, to complement the DKIST \citep{rast2021} data and to determine whether the chromospheric events exhibit detectable signatures in the SDO/AIA UV (1600 \AA) and EUV (mainly in 304, 171, and 193 \AA) channels. AIA gives high spatial resolution images (0.6\arcsec\ per pixel) with temporal cadence of 12 s in seven EUV channels \citep{lem12}. 

We use line-of-sight (LOS) magnetograms from SDO/Helioseismic and Magnetic Imager (HMI; \citealt{scherrer12,schou12}). The HMI magnetograms were obtained with a temporal cadence of 45 seconds, a pixel size of 0.\arcsec5 (360 km), and a noise level of approximately $\pm$7 Gauss \citep{couvidat16}.  Both AIA and HMI datasets were downloaded from the JSOC website\footnote{http://jsoc.stanford.edu/ajax/exportdata.html}.

To enable a direct comparison between the ViSP raster scans and the HMI LOS magnetograms, the ViSP Stokes profiles were first inverted. A cross-correlation between the ViSP magnetogram at each scan position and the nearest-in-time HMI magnetogram  was then performed to establish accurate co-alignment. Based on this registration, a co-temporal HMI magnetogram was constructed over the full ViSP four-pointing mosaic (Figure 4(a) in \citealt{santos23}). The HMI map was generated by extracting slices from HMI images at coordinates corresponding to the ViSP slit positions at their respective acquisition times. The full HMI field of view was subsequently reconstructed using the ViSP 6302 raster pixel scale and cropped to the region of interest defined by the network regions in the H$\beta$ images. In this work, we use the constructed HMI map in order to do the one to one comparison with the ViSP magnetogram. Further details on the data processing can be found in  \cite{santos23}, and \cite{kuridze24}.

The resulting datasets from ViSP and HMI show strong overall agreement; however, the ViSP magnetogram reveals substantially finer spatial structure and greater magnetic field strengths than HMI, as expected (Figure \ref{fig2}). These differences are particularly pronounced in the minority-polarity flux, reflecting both the higher spatial resolution and the enhanced sensitivity of ViSP to weak magnetic flux. The noise level of the ViSP magnetograms is approximately $\pm$2 G \citep{santos23} in contrast to $\pm$7 G for HMI.

We use the DKIST/ViSP raster scan magnetogram (see details in \citealt{santos23}), which include time information for each slit position, to identify regions  hosting minority-polarity flux at the bases of small-scale features. The high-sensitivity ViSP raster magnetograms are first analyzed to locate sites exhibiting mixed-polarity magnetic structure associated with these events. We then examine the corresponding HMI magnetograms using the time information from the ViSP scan, to determine whether the same mixed-polarity signatures are detectable within the sensitivity and spatial resolution limits of HMI. 

The lifetime of each event is defined as the time interval between the onset of brightening (time of the first frame showing the feature) and the disappearance of the small-scale feature (time of the last frame showing the feature).  The feature's length is measured from its visible base to its visible tip at the opposite end at the time of its maximum extent. The feature's width is defined as the maximum lateral extent of the feature during peak brightness. All of these measurements are done using H$\beta$ images.

\begin{figure*}
	\centering
	\includegraphics[width=\linewidth]{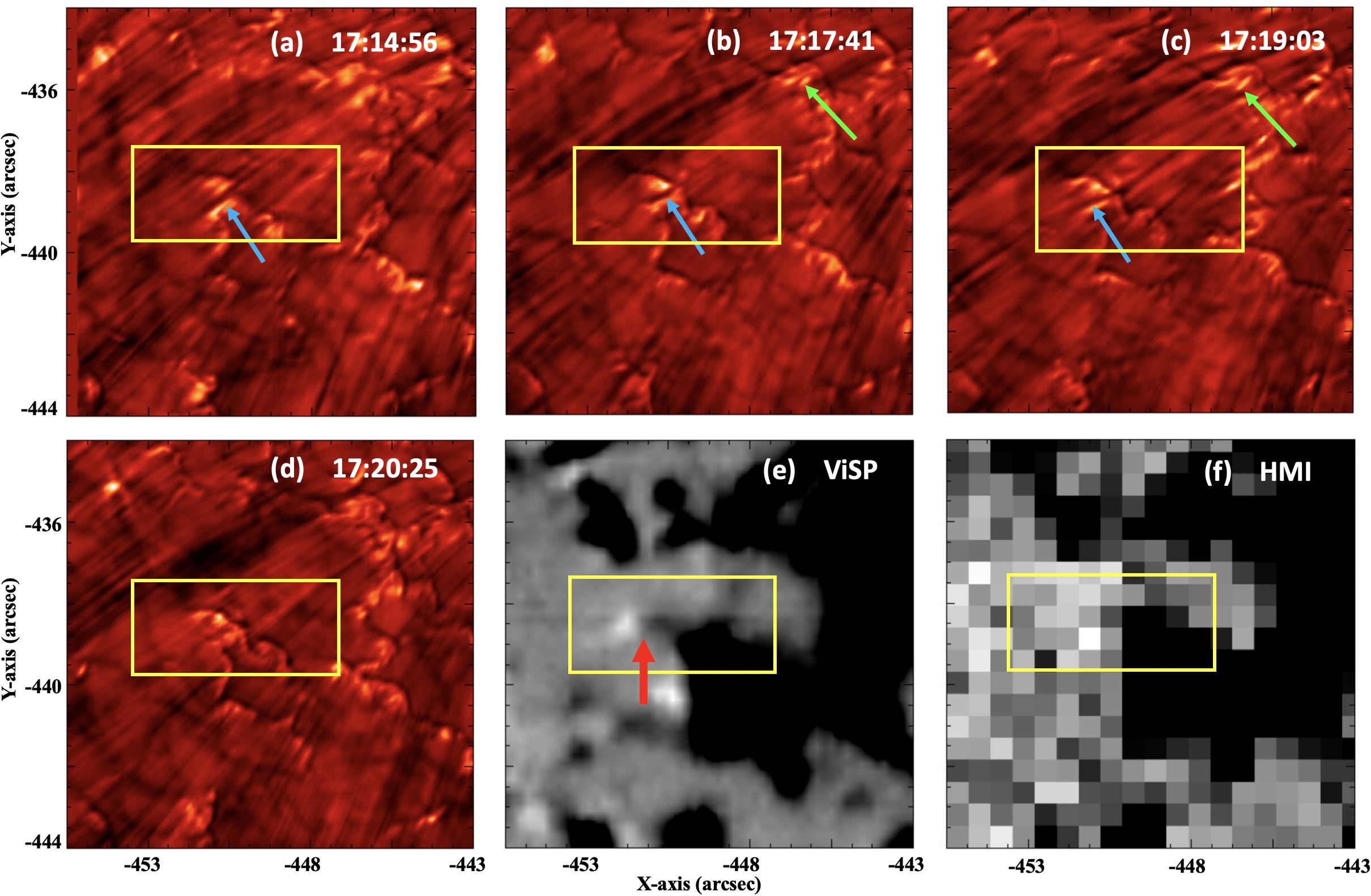}
	\caption{Another example H$\beta$ spicule and another example H$\beta$ microflash.  Panels (a–d) show VBI H$\beta$ images  of a  spicule (blue arrows) and an H$\beta$ microflash (green arrows). Panels (e) and (f) are the co-temporal ViSP magnetogram and constructed HMI magnetogram of the same region.  Magnetic field values are displayed within a range of $\pm$30 G. The yellow box centers on the spicule.  The red arrow points to an edge of minority-polarity (positive) flux near the spicule's base. An animation of panels (a) - (d) is available. This animation shows a larger field of view and runs from June 3rd, 2022 at 17:10:50 to 17:35:30. The animation real-time duration is 1.9 seconds. The same yellow box and blue and green arrows are used in the animation.
		\label{fig5}}
\end{figure*}

\begin{figure*}
	\centering
	\includegraphics[width=\linewidth]{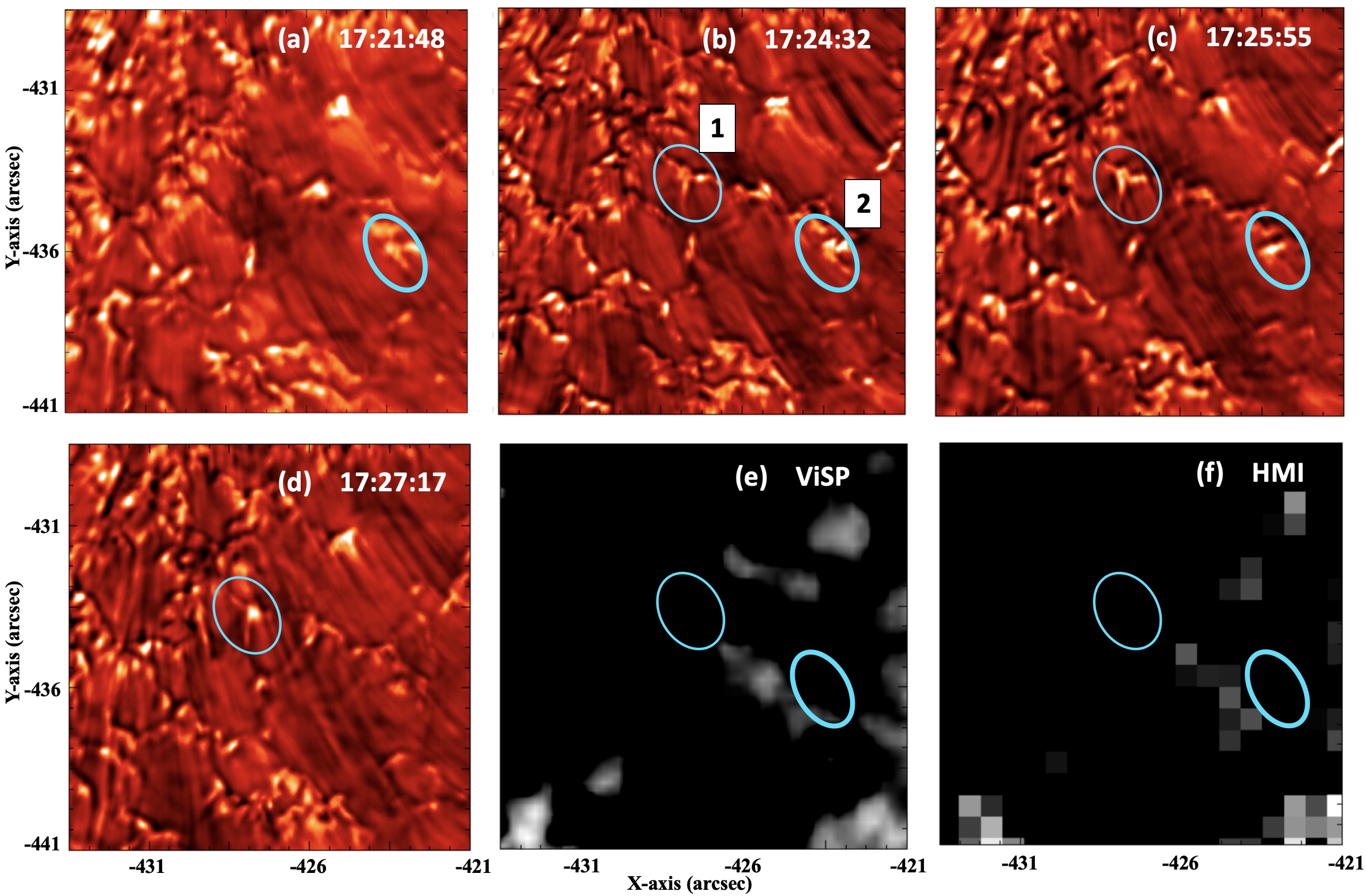}
	\caption{Two more examples of small-scale jets. Panels (a–d) show VBI H$\beta$ images showing two small-scale jets (blue oval). The small-scale jets are labeled as `1' and `2'. Panels (e) and (f) present the corresponding ViSP and HMI magnetograms of the same region.  Magnetic field values are displayed within a range of $\pm$30 G. An animation of panels (a) - (d) is available. This animation shows a larger field of view and runs from June 3rd, 2022 at 17:10:50 to 17:35:30. The animation real-time duration is 1.9 seconds. The same blue ovals are used in the animation. 
		\label{fig6}}
\end{figure*}

\section{Results} \label{sec:res}
\subsection{Overview}

Figure \ref{fig1} provides an overview of the DKIST pointing for the VBI and ViSP observations analyzed in this study. We focus on a single VBI pointing (the pointing having the H$\beta$ images that look most sharp or least blurry) and its corresponding LOS ViSP raster scans and constructed HMI magnetogram that is matched to the spatiotemporal locations of the ViSP slit. The AIA 171 \AA\ image (Figure \ref{fig1}a) indicates that the DKIST field of view is in a quiet-Sun region, where the magnetic network is predominantly of negative polarity. Coronal plumes are observed to originate from these network flux concentrations observed by DKIST. The black box outlines the region examined in detail in the subsequent panels.

Figure~\ref{fig1}b shows fibril-like structures extending outward from the plage region identified in Figure~\ref{fig1}c, consistent with the observations reported by \citet{santos23}. In this work, we study small-scale chromospheric features, (H$\beta$ spicules, small-scale jets, and H$\beta$ microflashes), together with a coronal plume, and examine their LOS photospheric magnetic flux using ViSP and HMI observations. Both HMI  (6302 \AA) and ViSP magnetograms  (8542 \AA) show a lane of negative-polarity magnetic flux across the region. Most of the small-scale features originate in regions of negative-polarity magnetic flux, in both photospheric and chromospheric magnetograms. We primarily use photospheric magnetic field measurements from SDO/HMI 6173~\AA\ and DKIST/ViSP 6302~\AA\ in our analysis.

\subsection{Photospheric magnetic fluxes: ViSP and HMI}\label{sec:magfield}

Before assessing the presence of minority-polarity flux at the base of the chromospheric events, we first compare magnetic flux histograms from the ViSP and HMI magnetograms. The histograms are from the field of view in Figures \ref{fig1}d and \ref{fig1}e. The resulting histograms (Figure \ref{fig2}) show striking general agreement but indicate that the ViSP magnetogram is significantly more sensitive to weak magnetic field, exhibiting a clear enhancement at low flux values compared to HMI.

Although the region is dominated by negative-polarity flux, ViSP reveals a substantially larger population of weak positive-polarity pixels than HMI, as highlighted in the inset of Figure \ref{fig2} specifically for 10 G or less. This difference reflects the presents of sub-HMI-pixel positive-polarity flux elements that are below the detection threshold or spatial resolution of HMI. Consequently, ViSP provides more evidence than HMI for  the presence of weak mixed-polarity magnetic fields at the base of several chromospheric events, including cases where HMI shows no above-noise positive field.

However, there are also instances in which neither ViSP nor HMI reveals any detectable opposite-polarity field within the analyzed regions where dynamic events occur (Figures \ref{fig1}d and \ref{fig1}e). The difference between the two instruments is also directly evident in the magnetograms shown in Figures \ref{fig1}d and \ref{fig1}e, particularly in the resolution of fine-scale magnetic structure.

Furthermore, we constructed a density plot comparing the ViSP and HMI magnetic field measurements for the field of view shown in Figures \ref{fig1}d and \ref{fig1}e. The resulting density distribution is presented in Figure \ref{fig2a}. The distribution shows that the HMI magnetic flux strength measurements tend to be lower than the corresponding ViSP measurements, indicating that HMI underestimates the magnetic flux strength relative to ViSP. The black dashed line represents the 1:1 reference line. The distribution is not symmetric about this line, with a larger concentration of pixels toward lower HMI field values, further demonstrating the systematic difference between the two instruments. These results are consistent with the findings of \cite{santos23}, although their analysis was performed over a larger FOV.

\subsection{Chromospheric small-scale events: H$\beta$ spicules, small-scale jets, and H$\beta$ microflashes}\label{sec:events}

In this section, we examine small-scale chromospheric events observed in DKIST H$\beta$ images, including H$\beta$ spicules, H$\beta$ microflashes, and small-scale jet-like features. These events are not clearly detected in the SDO/AIA UV and EUV channels (e.g., 1600~\AA, 304~\AA, and 171~\AA), suggesting that their signatures are primarily confined to chromospheric layers, although differences in spatial resolution between DKIST and AIA may also contribute to their limited visibility in AIA images. Most of these events originate within or near magnetic network lanes, indicating a close association with concentrated photospheric magnetic field. We define H$\beta$ spicules \citep{krat1971,alissandrakis1973,bose23} as relatively dark, elongated features whose widths and lengths are comparable to those of chromospheric spicules \citep{pereira12,tian14}. We call smaller chromospheric jets small-scale jets; these have a bright base and have narrow spires that exhibit both bright and dark structures.  H$\beta$  microflashes are defined as still-smaller transient brightenings that persist for more than two consecutive VBI images. All measured properties of the small-scale H$\beta$ chromospheric events are listed in Table~\ref{tab:list}.

\textbf{H$\beta$ spicules:} Figure \ref{fig3} presents an example of a spicule (located inside the white dashed  box). 
Initially, the spicule appears compact; as time progresses, it elongates and broadens. The blue arrow points to the spicule  spire. This jet originates within a negative-polarity magnetic flux patch where no opposite-polarity flux is detectable, neither in the ViSP nor in the HMI magnetograms. The lifetime of this feature is 4.46 minutes. Extending outside from the plage region, several fibril-like structures are also visible \citep{santos23}. The maximum length of the spicule  is 3740 $\pm$ 120 km and width is 890 $\pm$ 170 km. The plane of sky speed of H$\beta$ spicule is 13.5 $\pm$ 3.0 \kms.

A second example of a H$\beta$ spicule is shown in Figure \ref{fig4}. This spicule is smaller than the previously described event and has a lifetime of 4.0 minutes. The H$\beta$ images show a localized brightening (pointed to by the yellow arrow) adjacent to the dark spire, resembling the jet-like structure or surge-like structure reported by \cite{panesar19,tiwari19,sterling20} using Hi-C 172 \AA\ data. The brightening fades once the spire disappears, suggesting a close temporal and spatial connection. Its maximum length is 3200 $\pm$ 230 km and width is 220 $\pm$ 80 km. The speed of H$\beta$ spicule is 15.5  $\pm$ 13 \kms.

Importantly, this spicule originates at the edge of a magnetic network flux concentration, where small-scale opposite-polarity field is detected in the ViSP magnetogram (Figure \ref{fig4}e, yellow box) but remains undetected in HMI's lower spatial resolution and higher magnetic-flux noise level. 

Figure~\ref{fig5} shows a third example of a H$\beta$ spicule that is similar to the second example. The dark spicule extends southwest, accompanied by an adjacent bright streak (indicated by the blue arrows). The ViSP magnetogram reveals that the spicule  is rooted in a tiny positive-polarity magnetic flux element (red arrow in Figure~\ref{fig5}e), whereas the HMI magnetogram shows no clear evidence of positive flux at the jet base. The morphology of this H$\beta$ spicule closely resembles the H$\alpha$ spicular activity reported in Figure~6 of \cite{jeongwoo24} and in \cite{jeongwoo26}. The spicule reaches a maximum length of $650 \pm 45$ km and a maximum width of $144 \pm 12$ km, with a lifetime of 4.47 minutes. Its projected speed is $6.5 \pm 4.0 $ \kms.

Based on their observed lifetimes, lengths, and widths, these H$\beta$ spicules are likely the on-disk counterparts of spicules \citep{pontieu07-spicules,pereira12,dong-dkist-spicule2026}. Their properties are consistent with the network jets reported by \cite{tian14}; however, the absence of IRIS 1330~\AA\ observations in our dataset prevents a direct one-to-one comparison of their properties. Moreover, associations between spicules and mixed-polarity magnetic regions have also been reported using BBSO observations \citep{samanta2019,jeongwoo24}, suggesting that small-scale magnetic flux cancelation may play an important role in causing chromospheric jet phenomena.

The ability of ViSP to reveal such fine-scale minority-polarity magnetic features is scientifically crucial, as it enables us to directly test whether magnetic reconnection and corresponding flux cancelation — processes hidden in lower-resolution data - are responsible for preparing and triggering such H$\beta$ spicules. Identifying and quantifying such small-scale minority-polarity flux with ViSP  constrains and allows magnetic initiation mechanisms for spicule-like and jet-like events and other fine-scale energy release events (e.g. small-scale brightenings; \citealt{tiwari19,berghmans2021,panesar2021,tiwari22}) in the lower solar atmosphere.

\textbf{H$\beta$ microflashes:} We call the smallest brightenings that appear in DKIST H$\beta$ images of network regions  `H$\beta$ microflashes.' These events are small-scale, short-lived brightenings that appear as compact, slightly elongated structures within predominantly unipolar magnetic regions.   Two examples are shown in Figure \ref{fig3} (green arrows). The microflash on the right-hand side has a length of 480 $\pm$ 40 km and a width of 200 $\pm$ 30  km, with a lifetime of 6.9 minutes, and t moves with an average speed of 2.9 $\pm$ 0.2 \kms. The left-hand microflash has a maximum length of 320 $\pm$ 8 km and a width of 240  $\pm$ 20 km, and persists for approximately 4.5 minutes. It lengthens with an average speed of 3.5 $\pm$ 0.9 \kms.

Figure \ref{fig5} presents another example of an H$\beta$ microflash. This event has a maximum length of 295 $\pm$ 60 km and a width of 108 $\pm$ 10 km, with a lifetime of 1.6 minutes. It moves with an average speed of 3.0  $\pm$ 0.9 \kms. We note that these are small-scale brightenings that often change shape and are sometimes visible in only two consecutive frames.

The H$\beta$ microflashes may be chromospheric counterparts of the EUV microflashes reported by \citet{panesar2026}, which occur at the bases of coronal plumes rooted in unipolar network magnetic flux concentrations. Confirming that, however, will require future coordinated observations between DKIST and Solar Orbiter's Extreme Ultraviolet Imager (EUI). H$\beta$ microflashes are possibly signatures of sudden magnetic reconnection in the chromosphere that may contribute to chromospheric and coronal heating through wave generation and dissipation. Given the relatively transparent nature of H$\beta$, some of the features observed in these broadband images may be photospheric. Future ViSP observations will be valuable for constraining the formation and origin of these microflashes.

\textbf{Small-scale Jets:} Figure \ref{fig4} shows an example of a small-scale jet (inside the blue oval). The jet begins with a compact base brightening that becomes progressively brighter and wider with time, resembling the base brightenings commonly observed in coronal jets \citep{sterling15,panesar16b,moore18}. The eruption then develops into a faint, narrow spire extending outward from the base and persists for approximately 8.1 minutes. The jet achieves a length of 630 $\pm$ 49 km and a width of 490 $\pm$ 75 km. The jet lengthens  with an average speed of 6.0 $\pm$ 4.3 \kms.

HMI magnetograms indicate that the jet originates from an apparently unipolar magnetic flux patch with no clearly detectable opposite polarity at its base, implying the absence of a magnetic neutral line. In contrast, the higher-resolution ViSP magnetogram reveal localized flux dips surrounding the jet site (white contours in Figure \ref{fig4}e near the blue oval), suggesting the possible presence of unresolved opposite-polarity magnetic flux near the jet base. Although similar signatures may also be present in the HMI data, they are much less distinct and difficult to identify without guidance from the ViSP observations.

Figure \ref{fig6} presents two more examples of small-scale jets. Jet-1 is morphologically similar to the jet shown in Figure \ref{fig4}, exhibiting a brightened base region and a narrow, faint elongated spire (Figure \ref{fig6}c). The jet has a length of 660 $\pm$ 70 km and a width of 385 $\pm$ 55 km, with a lifetime of 2.8 minutes. Jet-1 lengthens  with an average speed of 3.8 $\pm$ 1.1 \kms. Jet-2 in Figure \ref{fig6} displays a comparatively fainter spire and weaker base brightening. Its length is 575 $\pm$ 75 km, its width is 275 $\pm$ 55 km, and its lifetime is approximately 4 minutes. It lengthens  with an average speed of 2.8 $\pm$ 0.1 \kms.

Both jets originate from a negative-polarity magnetic flux patch, where no clear opposite-polarity flux is directly visible. However, the ViSP observations again reveal localized dips in the magnetic flux near the jet base regions, suggesting the possible presence of unresolved or hidden opposite-polarity flux at the jet footpoints. Similar signatures are also visible in the HMI magnetograms, although they are less distinct than those observed in the ViSP data.

If such flux dips are commonly observed at the base of small-scale jets, that would supportive of those jets originating from canceling and erupting bipolar magnetic fields, analogous to larger-scale coronal jets. Similar flux dips at the bases of jet-like events have also been reported in HMI observations of jets seen by Hi-C 2.1 \citep{panesar19}. Conversely, if these dips are rarely detected at the bases of small-scale chromospheric jets, that would suggest that at least some of these events are driven by a different mechanism, such as wave-related processes \citep[e.g.,][]{sterling00b,pontieu04}.

\floattable
\begin{center}
	\begin{table*}[ht]
\setlength{\tabcolsep}{3pt} \caption{Properties of H$\beta$ Chromospheric Small-scale Events \label{tab:list}}
\renewcommand{\arraystretch}{1.0}\begin{tabular}{c*{9}{c}}
			\noalign{\smallskip}\tableline\tableline \noalign{\smallskip}
			
			Event  &  Time  & lifetimes  & Spire length & Base width  & Speeds & Magnetic & Magnetic & Kinetic & Thermal \\
			no.  & (UT)  & (min) & (km)   & (km) &  (\kms)  & Setting (ViSP) & Energy (erg) & Energy (erg) & Energy (erg)\\
			\noalign{\smallskip}\hline \noalign{\smallskip}
			H$\beta$ spicules \\ \noalign{\smallskip}\hline \noalign{\smallskip}
			1 (Fig. \ref{fig3}) &  17:25:55  & 4.46 & 3740 $\pm$ 120  & 890 $\pm$ 170  & 13.5 $\pm$ 3.0 & Unipolar  &  - &  - &- \\ 
2 (Fig. \ref{fig3}) &  17:20:25 & 4.0 & 3200 $\pm$ 230  & 220 $\pm$ 80  & 15.5  $\pm$ 13  & Mixed Polarity & - &  - &- \\ 
3 (Fig. \ref{fig5}) &  17:14:56 & 4.47 & 650 $\pm$ 45  & 144 $\pm$ 12  &  6.5 $\pm$ 4.0 & Mixed Polarity  &  - &  - &- \\ 
average$\pm$1$\sigma$$_{ave}$  &  & 4.3$\pm$ 0.25 & 2600$\pm$ 1600 & 420$\pm$ 400& 12 $\pm$ 4.7 &  & 1.5 $\times$ 10$^{26}$ & 4.4 $\times$ 10$^{24}$ & 7.5  $\times$ 10$^{24}$   \\

			\noalign{\smallskip}\tableline\tableline \noalign{\smallskip}
			H$\beta$  Microflashes\\
			\noalign{\smallskip}\hline \noalign{\smallskip}
1 (Fig. \ref{fig3}) &  17:25:33 & 6.9 & 480 $\pm$ 40  & 200 $\pm$ 30  & 3.0  $\pm$ 0.2 & Unipolar &  - &  - &- \\ 
2 (Fig. \ref{fig3})  &  17:25:55& 4.5 & 320 $\pm$ 8  & 240 $\pm$ 20  & 3.5  $\pm$ 0.9 & Unipolar &  - &  - &-  \\ 
3 (Fig. \ref{fig5}) &  17:17:41 & 1.6  & 295 $\pm$ 60  & 108 $\pm$ 10  & 3.0  $\pm$ 0.9 & Unipolar &  -  &  - &- \\   
			average$\pm$1$\sigma$$_{ave}$  &  & 4.3 $\pm$ 2.6 & 365 $\pm$ 100 & 180 $\pm$ 65& 3.1 $\pm$ 0.2  &  & 4.8 $\times$ 10$^{24}$ & 1.0 $\times$ 10$^{22}$ & 2.5  $\times$ 10$^{23}$   \\

			\noalign{\smallskip}\tableline\tableline \noalign{\smallskip}
			Small-scale Jets\\
			\noalign{\smallskip}\hline \noalign{\smallskip}
1  (Fig. \ref{fig4}) &  17:20:25& 8.0 & 630 $\pm$ 50  & 490 $\pm$ 75  & 6.0  $\pm$ 4.3 & Near Dips & - &  - &-  \\ 
2 (Fig. \ref{fig6}) &  17:21:48 & 2.8  & 660 $\pm$ 70  & 385 $\pm$ 55  & 4.0 $\pm$ 1.1 & Near Dips &  - &  - &-  \\ 
3 (Fig. \ref{fig6}) &  17:24:32 & 4.0  & 575 $\pm$ 75  & 275 $\pm$ 55  & 3.0 $\pm$ 0.1  & Near Dips &  - &  - &- \\   
			average$\pm$1$\sigma$$_{ave}$  &  & 5.0 $\pm$ 3.0 & 620 $\pm$ 40 & 385 $\pm$ 100 & 4.4 $\pm$ 1.5 & - & 2.8 $\times$ 10$^{25}$ &  1.2 $\times$ 10$^{23}$ & 1.5 $\times$ 10$^{24}$  \\

			\noalign{\smallskip}\tableline\tableline \noalign{\smallskip}
		\end{tabular}
\end{table*}		
\end{center}

\subsubsection{Energy estimates}

We estimate an upper limit to the magnetic energy released during the H$\beta$ spicules, small-scale jets, and H$\beta$ microflashes using the expression $B^{2}V/8\pi$, where $B$ is the magnetic field strength and $V$ is the volume of the pre-eruption magnetic field. For these calculations, on a conservative side we assume a magnetic field strength of 100 G. 	To estimate the volume ($V$), we assume a cylindrical geometry, $V=\pi r^{2}l$, where $l$ is the jet length and the observed jet width is taken to be the cylinder diameter ($2r$) \citep{sterling17}.  The estimated magnetic energies for the H$\beta$ spicules, small-scale jets, and H$\beta$ microflashes are approximately $1.5 \times 10^{26}$ erg, $2.8 \times 10^{25}$ erg, and $4.8 \times 10^{24}$ erg, respectively.
	
The estimated magnetic energies are comparable to those reported for some campfires \citep{panesar2021}, coronal hole jets \citep{pucci13}, coronal bright points \citep{priest94}, and bright dots \citep{tiwari19,tiwari22}. However, they are one or more orders of magnitude lower than the estimated magnetic energies of active-region jets \citep{sterling17,panesar2025} and quiet-region jets \citep{panesar16b, panesar2025}.

We also estimate the thermal and kinetic energies of the H$\beta$ spicules, small-scale jets, and H$\beta$ microflashes. The thermal energy is calculated as $E_{\rm th} = 1.5N_{e}k_{B}TV$, and the kinetic energy as $E_{\rm kin} = 0.5N_{e}m_{p}Vv^{2}$, where $N_{e}$ is the electron number density, $k_{B}$ is the Boltzmann constant, $T$ is the plasma temperature, $V$ is the emitting volume, $m_{p}$ is the proton mass, and $v$ is the jet speed listed in Table~\ref*{tab:list}. Adopting representative values of $v = 12 $ km s$^{-1}$, $N_{e} = 10^{13}$ cm$^{-3}$ \citep{judge1998}, and $T = 10^{4}$ K, we estimate the thermal and kinetic energies of the H$\beta$ spicules to be approximately $7.5 \times 10^{24}$ erg and $4.4 \times 10^{24}$ erg, respectively. Similarly, the thermal and kinetic energies of the small-scale jets are estimated to be approximately $1.5 \times 10^{24}$ erg and $1.2 \times 10^{23}$ erg, respectively. For the H$\beta$ microflashes, the estimated thermal and kinetic energies are approximately $2.5\times10^{23}$ erg and $1.0\times10^{22}$ erg, respectively, about one order of magnitude lower than those reported for EUV microflashes, which have thermal and kinetic energies of order $10^{24}$ erg and $10^{23}$ erg, respectively \citep{panesar2026}.
	
The estimated kinetic and thermal energies of our small-scale jets are approximately two orders of magnitude lower than those reported for EUV jets by \cite{jeongwoo25}. Their thermal energy is also about one order of magnitude lower than that of large penumbral jets, which have been reported to release thermal energies of order $10^{25}$ erg \citep{tiwari18}. In contrast, the estimated kinetic and thermal energies are comparable to those reported for coronal microjets \citep{hou21}, H$\alpha$ minifilaments \citep{jiashengwang24}, and quiet-region jets \citep{panesar2025}. These similarities indicate that small-scale chromospheric jets release energies comparable to those of other small-scale solar transients across different atmospheric environments. Their thermal energies, of order $10^{24}$ erg, are comparable to the characteristic energy of nanoflares ($\sim10^{24}$ erg) proposed by \cite{parker88}, suggesting that these small-scale jet-like events may contribute to transient heating of the chromosphere and low corona.

The estimated energies of our H$\beta$ spicules are comparable to those reported for chromospheric spicule-like events. Previous studies have shown that spicules can carry significant amounts of energy and mass into the upper atmosphere, with individual events requiring energies of order $10^{24}$--$10^{26}$ erg depending on their physical properties and assumed plasma parameters \citep{beckers68,pontieu09,kuridze15,samanta2019}. Such energetic chromospheric transients may contribute to coronal heating and solar-wind acceleration by transporting mass, momentum, and energy from the lower atmosphere into the upper solar atmosphere \citep{pontieu04,innes09,moore11,jeongwoo25,panesar2026}. The comparable energy range suggests that the H$\beta$ spicules observed here may represent a population of chromospheric jet phenomena related to spicules and may contribute to the mass and energy supply of the corona and solar wind.

\begin{figure}
	\centering
	\includegraphics[width=0.97\linewidth]{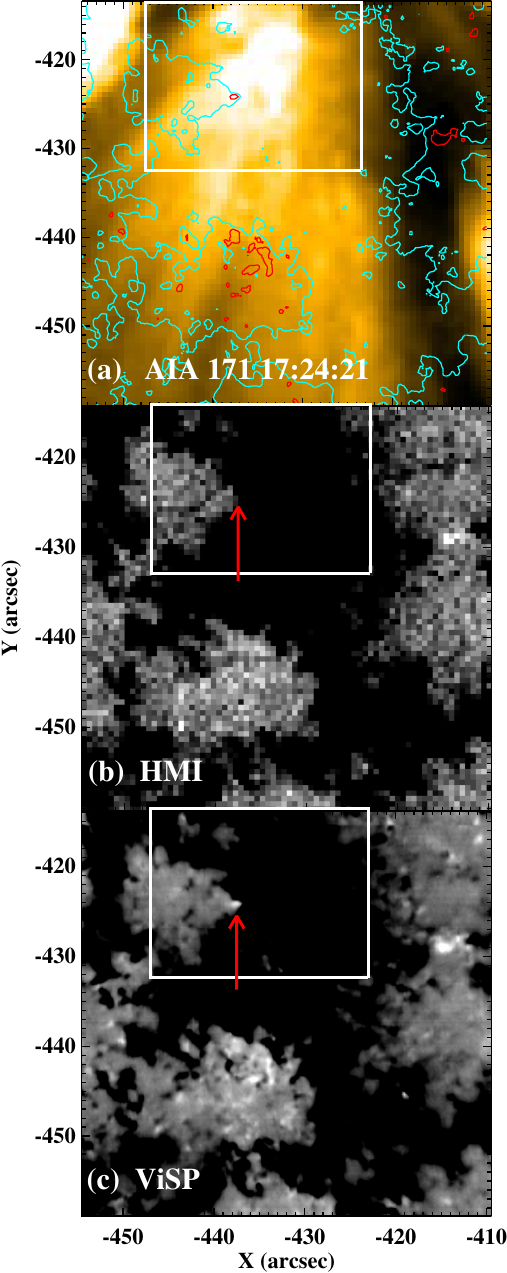}
	\caption{Coronal Plume and its magnetic flux. Panel (a) shows an AIA 171 \AA\ image of the plume. Panels (b) and (c), respectively, show the corresponding HMI and ViSP magnetogram of the same region.  In panel (a), ViSP contours, of levels $\pm$15 G are overlaid, where red and cyan contours outline positive and negative magnetic flux, respectively. The white box marks the FOV shown in Figures \ref{fig8}b,c. The red arrow points to the inclusion of positive-polarity flux at an edge of the negative flux at the base of the plume.
		\label{fig7}}
\end{figure}

\subsection{Coronal Plume and its Underlying Magnetic flux}\label{sec:plume}

Figure \ref{fig7}a shows an AIA 171 \AA\ image of a plume. The underlying photospheric magnetic flux of the plume is well captured in the DKIST/ViSP observations. The plume is rooted in a negative-polarity magnetic flux patch (Figures \ref{fig7}b,c). Using the ViSP magnetogram, we investigate whether mixed-polarity magnetic flux is present at the plume base. Previous studies have suggested that unresolved or weak mixed-polarity field may exist at plume footpoints \citep{wang-y-m2016,avallone18}, remaining undetected in  HMI observations. 

The plume emission in Figure \ref{fig7}a is primarily concentrated near the center of the negative-polarity magnetic flux at the plume's base (see contours in Figure \ref{fig7}a). The HMI and ViSP magnetograms are shown over the same field of view and displayed using an identical magnetic field scale. In the ViSP magnetogram, a small clump of positive-polarity flux is visible along the left edge of the dominant negative-polarity network patch (see red arrow in panel c), whereas no corresponding positive flux is detected  in the HMI constructed magnetogram in \ref{fig7}b - the red arrow is placed at the same location as the ViSP magnetogram in \ref{fig7}c. The presence of this positive-polarity field causes the boundary of the negative-polarity network in the ViSP observations to appear slightly westward relative to that seen in HMI. This suggests that magnetic-flux depressions/dips observed in HMI magnetograms may be considered as signature of unresolved opposite-polarity flux, consistent with the findings of \cite{panesar19}. 
Apart from this feature, we do not identify any clear opposite-polarity flux patches or clumps either near or along the boundary of the dominant negative-polarity network flux region. The ViSP magnetogram shows two depressions/dips in the negative-polarity flux distribution (near x=-443" and y=-415"), which are also present in the constructed HMI magnetogram. Although these dips appear slightly larger in the ViSP data compared to HMI, they do not exhibit any detectable positive-polarity flux concentrations. Future statistical observations with ViSP and the Visible Tunable Filter (VTF; \citealt{schmidt14}) on DKIST are needed to further investigate the presence of magnetic dips.

\begin{figure}
	\centering
	\includegraphics[width=\linewidth]{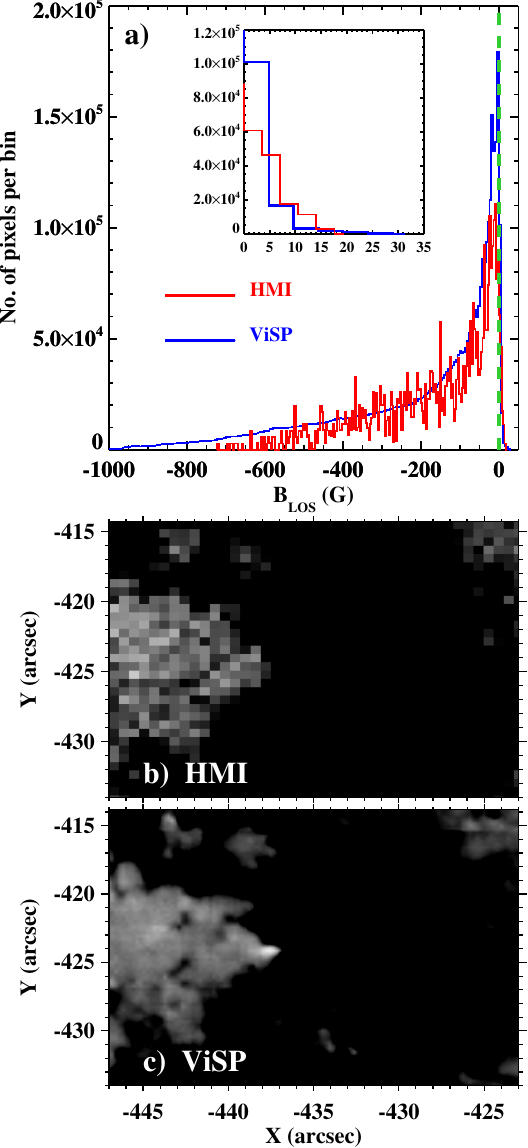}
	\caption{Positive and negative magnetic flux at the base of the coronal plume.  Panel (a) shows the histograms of the magnetic flux in the FOV of panels (b) and (c), comparing HMI with ViSP.  The green dashed vertical line marks 0 G.  The inset is for the positive flux.  The ViSP histograms are blue; the HMI histograms are red.  Panels (b) and (c) are the HMI and ViSP magnetograms in the white box in Figure \ref{fig7}.
		\label{fig8}}
\end{figure}

To examine the magnetic flux  at the base of the plume in greater detail, we selected a smaller field of view, indicated by the white box in Figure \ref{fig7}, and compared the HMI and ViSP magnetograms quantitatively. Figure \ref{fig8} presents the magnetic field histograms along with the zoomed-in field of view magnetogram from both HMI and ViSP. The comparison shows that the ViSP magnetogram detects stronger negative magnetic flux at the plume base, reaching values of approximately $-1000$ G, whereas the HMI magnetogram shows less-strong strongest flux of about $-700$ G. In addition, the ViSP data reveal relatively stronger positive-polarity magnetic fields, while in the HMI observations most of the positive flux values remain close to the noise level within the plume-base field of view (see inset in Figure \ref{fig8}a). Overall, the higher sensitivity and spatial resolution of ViSP reveal significantly stronger and more structured magnetic fields at the plume base compared to HMI.

Our observations indicate the possible presence of weak opposite-polarity flux at the plume base that remains unresolved in the HMI observations. Nevertheless, the overall magnetic configuration is consistent with previous plume studies that reported no significant opposite-polarity flux either at or near the plume footpoints \citep{panesar18b,avallone18,moore2023,weitz25,panesar2026}.

\section{Discussion}

Using the highest spatial resolution so far of DKIST observations together with SDO data, we investigated the dynamics and underlying magnetic field structure of sub-arcsecond chromospheric dynamic events. We also examined the differences between the magnetic field measurements obtained from ViSP and HMI. Our analysis shows that ViSP detects significantly stronger, as well as weaker, magnetic flux than HMI observations. Below we summarized our main findings: 

\begin{enumerate}

\item The comparison between the ViSP and HMI magnetograms demonstrates the importance of high-spatial-resolution and high-sensitivity spectropolarimetric observations for investigating the magnetic environment underlying chromospheric activity. The enhanced detection of weak magnetic flux by ViSP, particularly the larger population of small-scale positive-polarity elements embedded within the dominant negative-polarity region, suggests that mixed-polarity magnetic fields may be far more common than inferred from lower-resolution observations. These weak opposite-polarity flux patches are likely unresolved or below the sensitivity threshold of HMI, implying that some chromospheric and coronal events previously interpreted as originating from purely unipolar regions may in fact involve localized mixed-polarity magnetic interactions at much smaller spatial scales \citep{wang-y-m2016}.

Similar detections of small-scale opposite-polarity magnetic flux were also reported by \cite{manolis25} using LOS magnetograms from the Big Bear Solar Observatory's (BBSO) Goode Solar Telescope (GST). Compared with HMI, the GST/Near InfraRed Imaging Spectropolarimeter (NIRIS) observations reveal a wealth of small-scale magnetic features, including both unipolar and mixed-polarity fields. In contrast, the HMI magnetograms show only weak, diffuse, and predominantly unipolar magnetic features, leaving many of the small-scale magnetic structures unresolved \citep{abramenko20}. Similar conclusions have been drawn from comparisons between the Hinode Solar Optical Telescope/Spectro-Polarimeter (SOT/SP) and HMI observations. Owing to its higher spatial resolution and polarimetric sensitivity, Hinode/SP detects substantially stronger and more intricate magnetic fields than HMI, resulting in larger estimates of the magnetic flux and magnetic energy \citep[e.g.][]{thalmann13}.  Consistent with these findings, an active-region study by \cite{haimin-wang2017} demonstrated that NIRIS resolves photospheric magnetic structures in greater detail than both Hinode/SP and SDO/HMI, highlighting the importance of high-resolution observations for detecting the fine-scale complexity of photospheric magnetic fields.

In this study, we find that such hidden mixed-polarity structures could facilitate magnetic reconnection and contribute to the initiation of jets and plume-related activity. At the same time, the absence of detectable opposite-polarity flux in some events, even in the higher-sensitivity ViSP observations, suggests that not all chromospheric dynamics necessarily require observable mixed-polarity flux at photospheric levels. Such events may instead be driven by upward-propagating MHD waves, or by magnetic reconnection occurring within apparently unipolar magnetic field configurations \citep{panesar2026}.

In addition, these events might instead be driven by mechanisms operating higher in the atmosphere, unresolved magnetic structure below the current detection limit, or magnetic configurations associated with dips and highly inclined fields. Overall, our results emphasize that DKIST/ViSP observations provide critical insight into the fine-scale magnetic topology of the solar atmosphere and reveal magnetic complexity that is not accessible with HMI alone.

	\item Using VBI H$\beta$ observations, we find that both the H$\beta$ spicules and small-scale jets originate in and at  the edges of magnetic-network flux lanes, consistent with previously reported jetlets observed with IRIS \citep{panesar18b} and with Hi-C 2.1 \citep{panesar19}. Two of the three H$\beta$ spicules originate from locations where ViSP observations show signatures of opposite-polarity magnetic field, in agreement with the previous spicule studies. All three small-scale jets are rooted in regions exhibiting magnetic dips in the magnetograms and display compact base brightenings together with narrow, collimated spires. However, we do not discern an erupting flux rope in these events – a minifilament-carrying erupting flux rope usually is observed in larger jets and larger campfires.

If these chromospheric small-scale jets are miniature analogs of coronal jets, then the observed base brightenings may be interpreted as signatures of both internal and external magnetic reconnection. In this scenario, internal reconnection occurs between the legs of an erupting magnetic arcade that may contain a minifilament flux rope at its core, whereas external reconnection takes place between the erupting arcade and the surrounding large-scale magnetic field, following the scenario proposed by \cite{sterling15,panesar16b}. The jet spire would then be produced by external reconnection driven by the eruption of the minifilament-carrying magnetic field.

The average observed widths (420 $\pm$ 400 km), lengths (2600 $\pm$ 1600 km),  lifetimes (4.3 $\pm$ 0.25 minutes), and speeds (12 $\pm$ 4.7 \kms) of the H$\beta$ spicules are broadly consistent with previously reported chromospheric spicule properties, including widths of $\sim$400 km, lengths of 3000–6000 km, lifetimes of $\sim$3–10 minutes, and  speeds of  10--40 \kms\ \citep{pontieu07-spicules,sterling00b,pereira12}. 

The three small-scale chromospheric jets analyzed here exhibit an average spire length of 620 $\pm$ 40 km and an average base width of 385 $\pm$ 100 km. These lengths are substantially smaller than those reported for typical coronal jets, fine-scale EUI jets (6050 km; \citealt{panesar23}), and EUI microjets (7700 km; \citealt{hou21}). Similarly, their base widths are considerably narrower than those of campfires (1600 km; \citealt{panesar2021}) and fine-scale EUI jets (2200 km; \citealt{panesar23}). The average lifetime of the small-scale jets is 4.9 minutes, which is broadly comparable to the lifetimes reported for IRIS jetlets (3 minutes; \citealt{panesar18b}) and EUI jets (6.5 minutes; \citealt{panesar23}). These measured properties suggest that these chromospheric jets may represent scaled-down counterparts of a broader continuum of solar jet phenomena. However their average observed speed is low: 4.4 $\pm$ 1.5 \kms.

\item We find that H$\beta$ microflashes occur in unipolar magnetic flux regions, similar to the EUV microflashes recently reported by \cite{panesar2026}. However, unlike the EUV microflashes, the H$\beta$ microflashes analyzed here are located in non-plume regions. Their average lengths (365 $\pm$ 100 km) and widths (180 $\pm$ 65 km) are also smaller than those reported for EUV microflashes (860 $\pm$ 150 km and 770 $\pm$ 200 km, respectively). In contrast, their lifetimes (4.3 $\pm$ 2.6 minutes) are significantly longer than those of EUV microflashes (22 $\pm$ 20 seconds). Their average speed  is 3.1 $\pm$ 0.2 \kms.

It therefore remains unclear whether these chromospheric H$\beta$ microflashes represent lower-atmosphere counterparts of EUV microflashes or constitute a distinct class of small-scale transient brightenings. Addressing this question will require higher temporal cadence VBI observations, higher-cadence ViSP spectropolarimetry, and coordinated observations with the Solar Orbiter/EUI, which together will help clarify the physical nature of these small-scale events.

\item  Plumes are rooted in the most strongly unipolar magnetic flux regions within coronal holes. The ViSP magnetogram shows that there is a small opposite  positive-polarity flux patch (of $\pm$15 G, which is above the ViSP noise level of $\pm$2 G) present at the base of the plume that we examine. This opposite  positive-polarity flux patch is not observed by HMI. However, the HMI magnetogram shows a corresponding magnetic-field depression at the same location, suggesting the possible presence of unresolved opposite-polarity flux.
Coronal plumes  may form in mixed-polarity regions as suggested by \cite{wang-y-m2016}. However, emission may additionally originate well within the network concentration, where no visible opposite-polarity flux is detected \citep{panesar2026}. No jets or jet-like features are observed at the neutral-line location where the opposite-polarity flux patch is present \citep{raouafi14,panesar18b,panesar20b}. Instead, the plume exhibits only its characteristic outward plasma flows from unipolar network concentrations, forming a curtain-like haze \citep{wang-y-m2016}. Furthermore, the ViSP observations reveal stronger magnetic field (1000 G) strengths than HMI (700 G) at the base of the plume.

\end{enumerate}
	
Importantly, the DKIST/VBI high-resolution imaging provides precise information on the spatio-temporal evolution of chromospheric dynamic fine-scale structure and accompanied brightenings, allowing us to link magnetic drivers observed by ViSP with the chromospheric response captured by VBI. This combination enables, for the first time, a co-temporal, multi-parameter analysis of chromospheric jet formation -- revealing the interplay between magnetic flux evolution, plasma dynamics, and magnetic energy release at sub-arcsecond scales. Such coordinated ViSP–VBI observations will offer unprecedented resolution of the magnetic origins of spicule-like and jet-like activity in the lower solar atmosphere, providing new constraints for models of formation of those features, and models  of coronal heating and solar wind acceleration.  
	
Future high-resolution, high-cadence observations with DKIST/ViSP and VBI will provide stronger constraints on the magnetic-field structure and temporal evolution of small-scale solar features, enabling a more comprehensive investigation of their formation, evolution, and energetics. In particular, coordinated observations with the Solar Orbiter Polarimetric and Helioseismic Imager (PHI) and the EUI have the potential to directly detect previously unresolved opposite-polarity magnetic flux associated with these events, fundamentally improving our understanding of their origin and magnetic topology. Coordinated multi-instrument observations combining DKIST, IRIS, EUI, Solar-C, and MUSE will be essential for establishing the physical connection between chromospheric, transition-region, and coronal activity, thereby providing key constraints on the mechanisms of energy transfer and heating in the solar atmosphere.

\section{Conclusion}

We investigate the sub-arcsecond structure of the chromosphere using H$\beta$ and Fe I 6302 \AA\  observations obtained  with the largest solar optical telescope, DKIST, together with coordinated data from SDO/AIA and SDO/HMI. We find that ViSP spectropolarimetric measurements have substantially enhanced sensitivity to weak and small-scale magnetic flux compared to HMI, in particular for flux below 10 G. In multiple cases, ViSP detects clear mixed-polarity magnetic flux at or near the base of chromospheric events (e.g., H$\beta$ spicules) as well as at the base of coronal structures such as plumes, while HMI shows no corresponding opposite polarity magnetic field above 10 G.

We also identify events that appear to originate from nominally unipolar regions, where neither ViSP nor HMI detects opposite-polarity flux. In addition, several events are rooted in locations exhibiting magnetic dips, strongly suggesting the presence of unresolved mixed-polarity magnetic fields below the detection limits of current instrumentation.

These results demonstrate that a substantial fraction of small-scale chromospheric and coronal activity is rooted in magnetic structures that remain unresolved in current full-disk magnetograms such as HMI. They highlight the importance of high-spatial-resolution, high-sensitivity spectropolarimetric observations with DKIST for uncovering fine-scale magnetic flux arrangements.

\begin{acknowledgments}
We sincerely thank the anonymous referee for their helpful constructive comments and suggestions. We thank Kevin Reardon for performing the HMI–DKIST coalignment.
The research reported herein is crucially based on data collected with the Daniel K. Inouye Solar Telescope (DKIST), a facility of the National Solar Observatory (NSO). NSO is managed by the Association of Universities for Research in Astronomy, Inc., and is funded by the National Science Foundation. DKIST is located on land of spiritual and cultural significance to Native Hawaiian people. The use of this important site to further scientific knowledge is done so with appreciation and respect. NKP acknowledges support from NASA’s SDO/AIA (NNG04EA00C) grant, NASA's  HCSI (80NSSC25K7028) grant, and NASA’s HSR (80NSSC24K0258) grant.  SKT gratefully acknowledges support by NASA contract NNM07AA01C (Hinode). SKT, RLM, and NKP sincerely acknowledge support from  NSF AAG award (no. 2307505). SKT also acknowledges support from ARC-CREST (NASA Cooperative Agreement 80NSSC23M0230). 
ACS received support from NASA through the competed Heliophysics Internal Scientist Funding Model (HISFM) Program, and he benefited from discussions at the the International Space Science Institute(ISSI-BJ ID 24-604) team meeting on ”Small-scale eruptions in the Sun.”
We acknowledge the use of  \sdo/AIA/HMI data. AIA is an instrument onboard the Solar Dynamics Observatory, a mission for NASA’s Living With a Star program. This work has made use of NASA ADSABS and Solar Software.
\end{acknowledgments}

\bibliographystyle{aasjournal}

\end{document}